\documentclass[english,aps,pra,reprint,noshowpacs,superscriptaddress]{revtex4-1}   
\usepackage[T1]{fontenc}	
\usepackage[latin9]{inputenc}	
\usepackage{geometry}		
\usepackage{graphicx}
\usepackage[above,below]{placeins}	
\usepackage{times}
\usepackage{caption}
\usepackage{hyperref}  
\usepackage{float}
\usepackage{wrapfig}
\usepackage{subfig}
\makeatletter
\newcommand*{\balancecolsandclearpage}{%
  \close@column@grid
  \cleardoublepage
  \twocolumngrid
}
\makeatother
\usepackage{dcolumn}
\usepackage{bm}
\hypersetup{colorlinks=true,urlcolor=blue,citecolor=blue,linkcolor=blue}   
\begin{document}

\title{Virtual Reality as a Teaching Tool for Moon Phases and Beyond}
\author{J. H. Madden}
\affiliation{Astronomy and Space Science, Cornell University, Ithaca NY 14850}
\author{A. S. Won}
\author{J. P. Schuldt}
\author{B. Kim}
\author{S. Pandita}
\author{Y. Sun}
\author{T. J. Stone}
\affiliation{Communication, Cornell University, Ithaca NY 14850}
\author{N. G. Holmes}
\affiliation{LASSP, Cornell University, Ithaca NY 14850}

\begin{abstract}
A ball on a stick is a common and simple activity for teaching the phases of the Moon. This activity, like many others in physics and astronomy, gives students a perspective they otherwise could only imagine. For Moon phases, a third person view and control over time allows students to rapidly build a mental model that connects all the moving parts. Computer simulations of many traditional physics and astronomy activities provide new features, controls, or vantage points to enhance learning beyond a hands-on activity. Virtual reality provides the capabilities of computer simulations and embodied cognition experiences through a hands-on activity making it a natural step to improve learning. We recreated the traditional ball-and-stick moon phases activity in virtual reality and compared participant learning using this simulation with using traditional methods. We found a strong participant preference for VR relative to the traditional methods. However, we observed no difference across conditions in average levels of performance on a pre/post knowledge test.
\end{abstract}

\maketitle

\section{\label{sec:level0}Introduction}
The physics classroom today contains many forms of teaching designed to show students a different perspective on the material. Students can read a textbook, take notes in class, and observe demonstrations of the phenomenon being studied, among much else. The different routes of exposure are chosen to complement each other and provide new pathways to learn the material. 

A hands-on activity provides students with a personal, real, physical application of course content. This type of activity provides an embodied cognitive experience in which we make sense of the world through our body's interaction with it \cite{ANDERSON200391}. While embodied cognitive experiences enhance learning, the activities are often plagued by variability in fidelity, cognitive load, and sacrifices of realism that distract from the concepts being taught \cite{Finkelstein2005}. Implementing a heavily guided activity reduces these variables but also the opportunity for self-directed learning \cite{Weiman2015}. This is where the benefits of computer simulations and virtual reality step in to help. A computer simulation is able to reduce variability by providing a controlled experience and the realism can be programmed. A computer simulation that is designed with exploration in mind can also lend itself well to self-directed learning \cite{adams2008,Finkelstein2005}. 

Virtual Reality (VR) goes an extra step beyond the versatility and consistency of a computer simulation by also providing the embodied learning experience of a hands-on activity. VR brings other features that make it uniquely suited as a teaching tool in physics \cite{Bricken:1991:VRL:126640.126657,Perone2016}. 
It also allows for the use of inaccessible perspectives that are more immersive, and its ability to track students provides additional opportunities to monitor learning, engagement, and provide feedback in real time \cite{won}. 

A brief search through Physical Review - Physics Education Research (PER) and the PERC proceedings will show that very little research has been done in PER on the impacts of VR on learning. Recently, one study performed a controlled experiment comparing student learning and attitudes due to VR, video, and static images \cite{Smith2017}. They found no differences between conditions, except for a marginal improvement from VR for students with experience with video games. They also found significant participant preference towards VR.

A significant limitation to the study, however, was that the electrostatics concepts evaluated across conditions did not take advantage of the interactive and embodied cognitive affordances of VR. The students were not able to interact with the environment, control variables, or explore the concepts - features that are key for student learning from simulations \cite{Salehi2015,adams2008}.

In the present study, we aimed to test the impacts of a fully featured and interactive experience in VR. We compared participant learning and activity preference when engaging in equivalent activities in VR, a hands-on activity, and a desktop simulation. We found that, even with the interactive elements, participant learning was again equivalent across conditions, and that participants overwhelmingly preferred engaging in the VR condition.

\section{\label{sec:level1}Methods}
To carry out our study we created three Moon phase activities; a hands-on activity as close to traditional as possible, a computer based version, and a fully featured VR version. 

The traditional hands-on version of this activity involved a bright spotlight (the Sun) pointed at a participant (the Earth) who is holding a short stick with a ball on top (the Moon). If the participant holds the ball constantly at arm's length and spins around, this symbolic Sun-Earth-Moon system mimics the illumination pattern of the Moon phases so they can be viewed with the control of the participant \cite{tutorial,newbury}. 

In the desktop simulation version of the activity, participants saw realistic visualizations of the Sun, Earth, and Moon. Students used the keys on the computer keyboard to move forward or backwards in time. They could also change their viewing position by using the mouse and keyboard, for example to see a birds-eye-view of the 3-body system, or planar perspectives to more clearly see the shadows on the Moon. The time progression appropriately synced Earth's rotation, the Moon's orbit, and Earth's orbit.

The VR activity used the Oculus Rift head mounted display with two hand controllers running a simulation made using the Unity game engine. The structure of the VR activity was much the same as the desktop simulation, except that participants used the hand controllers to move forward or backwards in time. They could also use a cursor, virtually attached to their hand, to reach out and grab the Moon to drag it forwards or backwards in its orbit, in much the same way as the hands-on condition. Participants were initially placed in space above Earth's North pole, but could relocate to other viewing positions. 

The desktop simulation and VR conditions also had the opportunity for participants to transport to Earth's surface to observe rise and set times. They could fluidly transition between views to observe the rise and set of the Moon on Earth and the relative positions of the bodies at the same time. 

Participants were recruited from a pool of Cornell students and the instructions were kept as similar as possible for each condition. Each participant was randomly assigned to one of the three conditions. We used a pre-post model to measure participant learning and activity preference. The pre- and post-tests consisted of 14 questions each, drawn from existing assessments of student understanding of Moon phases \cite{2002AEdRv...1a..47H,lindell2002,lindell2005,slater2014}. Due to the short duration of the activity, the pre- and post-test items were not identical. The questions were, however, matched on content, such that each test contained a isomorphic set of questions. 


After completing the pre-test, the researcher provided verbal instructions on how to use the equipment for their condition. The language of the instructions were matched as closely as possible between conditions. The participant then completed an activity that used guiding questions modeled after common tutorials in astronomy \cite{tutorial,newbury}. We opted to use semi-structured activities as research has found that students can struggle to engage productively in unstructured computer-based activities \cite{Salehi2015,adams2008}. Once again, the guiding questions were equivalent between conditions.

Following the activity, the participant completed the post-test, which also included demographic and attitudinal questions. The participant was then shown the other two conditions and completed a short survey to indicate their preferred activity and to explain their reasoning.

There were 56 participants in the VR condition, 59 in the hands-on condition, and 57 in the desktop condition. All participants were students at Cornell University. 138 participants identified themselves as women, 31 as men, and 3 identified as other. The most common majors were social science, such as communication but there were also participants from biological science, engineering, physical science, and art. 

The results presented here, as gathered from the pre- and post-tests, include: participant performance by condition, participant performance by question, and activity preference. Video recordings were also collected for each participant during their time with the teaching activity and tracked movement data was collected for participants in the VR condition. The data have not yet been analyzed. 
 
\section{\label{sec:level2}Results}
\subsection{Total score comparisons}
\begin{figure}[H]
\centering
\includegraphics[width=\columnwidth]{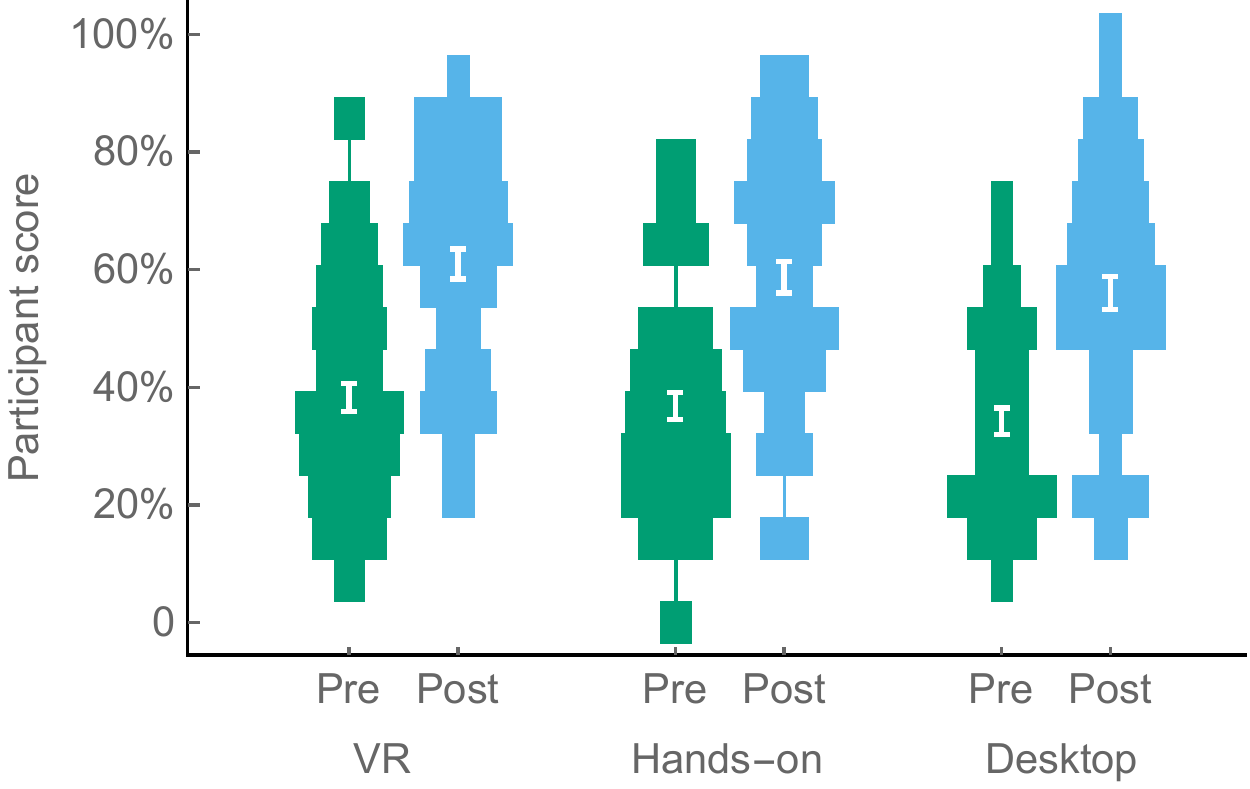}
\caption{\label{fig:Histograms} An overall view of pre- to post-test performance. Violin plots show how scores were distributed across conditions with green (left) being the pre-test and blue (right) being the post-test. Bins are 1 point wide. Average scores and standard error are indicated in white.}
\end{figure}

The average score on the pre-test was around 36\% in all three conditions, with no significant differences between conditions (Fig.\ref{fig:Histograms}). There were statistically significant increases in performance in all three conditions between pre- and post-test (\textit{p}<0.001). Because pre-test scores were statistically equivalent, we directly compared the post-test scores across conditions. The average score on the post-test was about 58\% in all three conditions, with no significant differences between conditions $(F(2,169)=0.86,$ $p=0.57)$. 

\subsection{Score comparison by topic}
\begin{figure}[H]
\centering
\includegraphics[width=\columnwidth]{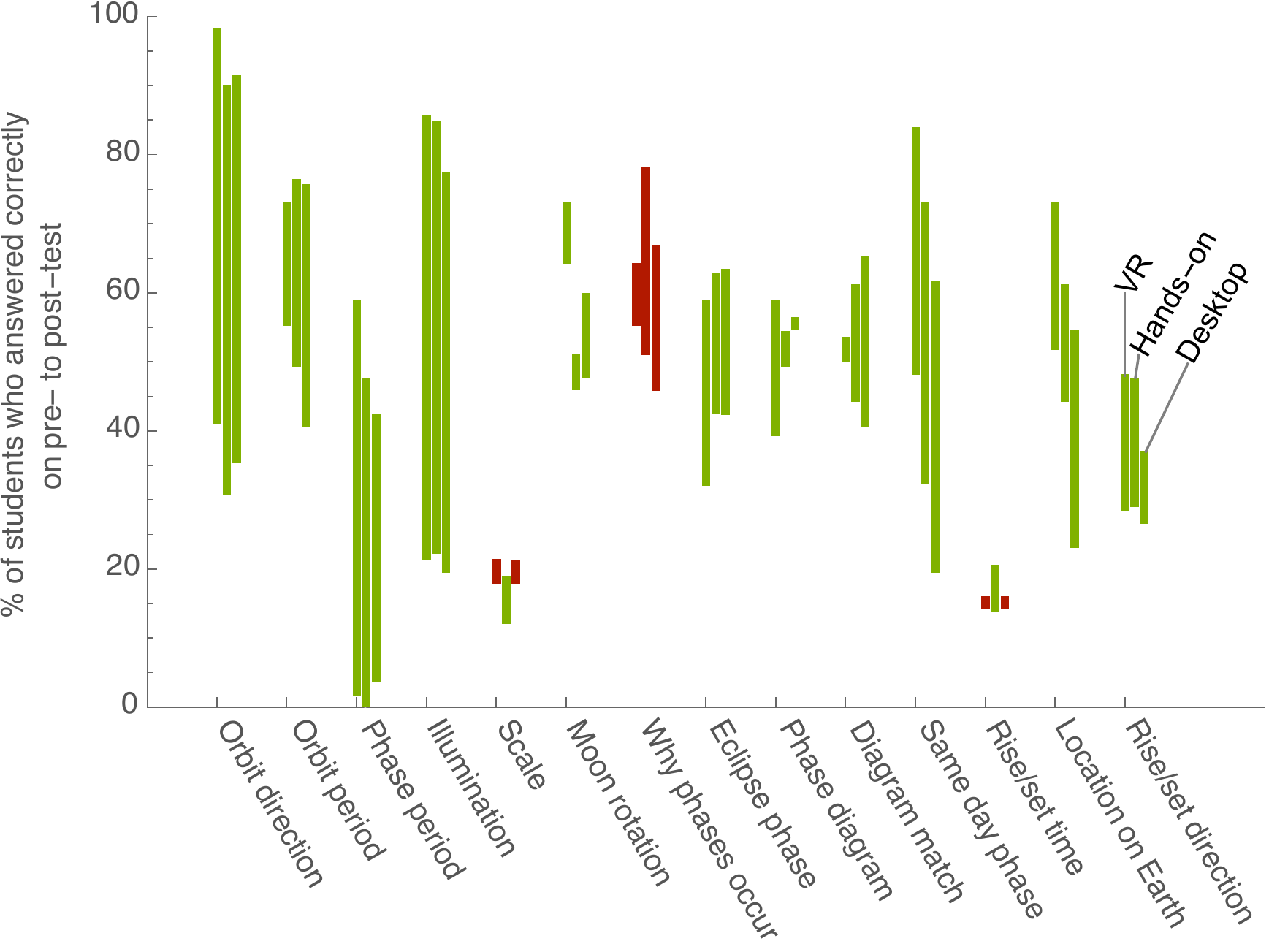}
\caption{\label{fig:prepostbyq} The difference in participant responses from the pre-test to the post-test broken down by question topic. The percent of correct responses on the pre-test and post test for that topic are connected by a colored bar. A green bar signifies improvement, with the higher number representing the post-test score. A red bar means there were fewer correct responses on the post-test, with the higher number representing pre-test score.}
\end{figure}

The three conditions provided different potential affordances for teaching particular content. For example, the orbit and rotation periods are not controlled in the hands-on condition, but are appropriately controlled in the desktop simulation and VR conditions. We would expect, therefore, that participants in the hands-on condition may perform worse on questions related to time scales. Upon being broken down by question the pre- to post-test scores show some variation between conditions but no statistically or practically significant differences (Fig. \ref{fig:prepostbyq}). A Fisher's exact test gives p-values greater than 0.1 for all pre/post scores by question. Some topic areas show gains over 40\% that are consistent across condition (orbit period, phase period, and illumination) while others show gains less than 10\%  (scale, Moon rotation, phase diagram, rise/set time), consistent with previous research on learning about Moon phases \cite{lindell2005,wilhelm2018}. Note the negative shifts in the ``Why phases occur" question. Upon investigation, the questions for this topic were not truly isomorphic. Though they referred to the same topic, they varied significantly in difficulty level.

\subsection{Participant preference}
\begin{figure}[!h]
\centering
\includegraphics[width=0.75\columnwidth]{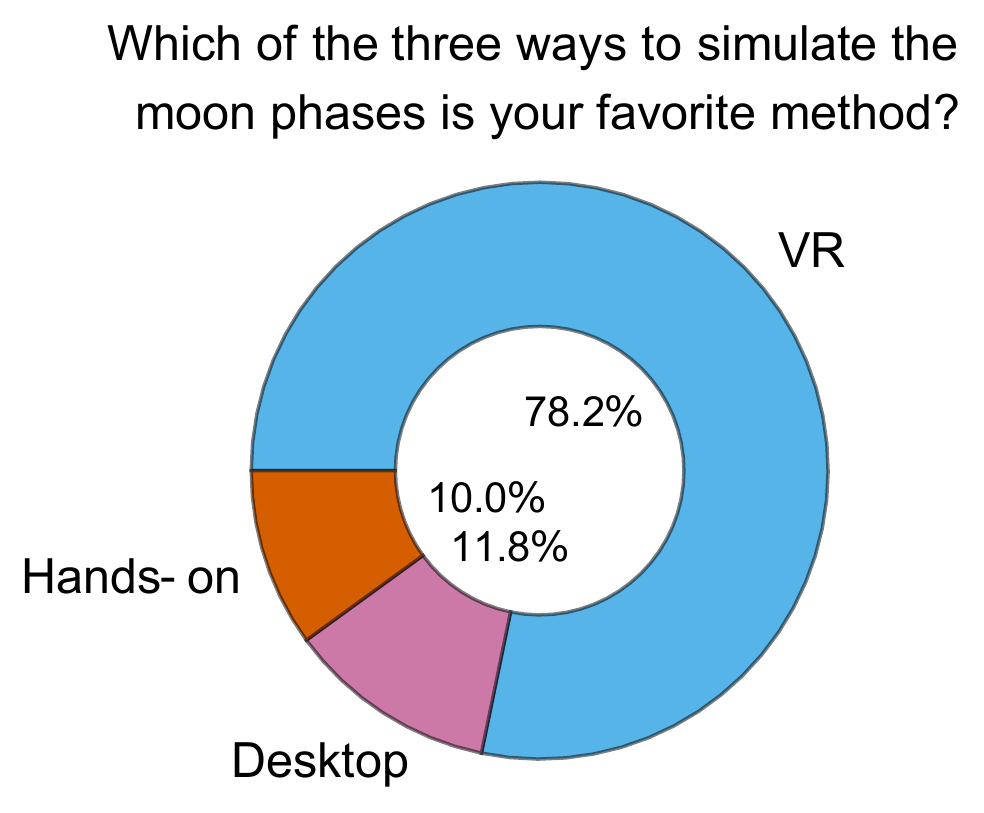}
\caption{\label{fig:Pref} Preferred choice after viewing each activity.}
\end{figure}

Despite these lack of differences between learning, we found that 78\% of participants preferred the VR activity (Fig. \ref{fig:Pref}) with no dependence on assigned condition. The participants who preferred VR generally found that seeing the full picture helped build their mental model. The main contributors to this were the ease of viewing different perspectives and having control over the system. For example, one participant wrote: 
\begin{quote}
``Having a overall space to see where everything is helps a lot. Even in class I still had a hard time understanding what they are talking about in concept. But I think I learned a lot in VR and being able to manipulate the environment on my own accord. It seems more engaging than the 2 other methods."
\end{quote}

If the participant did not prefer VR it was mostly because they felt uncomfortable with the overwhelming sensory input. They preferred either the hands-on or desktop conditions because they were more familiar. For example, a participant who preferred hands-on wrote:
\begin{quote}
``I really liked the virtual reality method. And it gave me more information than the other two methods, for instance, what time of day certain moon phases would rise and set. Nevertheless, it was almost too overwhelming and it was as if I was too excited to be in space to actually commit to learning the moon phases. With the hands-on demonstration. there was nothing to distract me. And, obviously, controlling the demonstration felt about as natural as possible." 
\end{quote}

A participant who preferred the desktop simulation wrote: 
\begin{quote}
``The VR was cool but since I'm very new to it I spent most of my time just trying to figure out how it worked--it was also tough to find where the sun and moon were at times because of how `large' the environment was. The desktop game was more familiar and easy-to-use for me. Personally."
\end{quote}

We can see that even participants who preferred other methods indicated that they liked the VR activity, but some participants were either overwhelmed by the controls or found it generally too distracting. As with computers, if the technology becomes more common, the issues regarding control and novelty may one day no longer be relevant. Currently however, efforts should be made to ensure familiarity before optimal efficiency is attained. 

\section{Discussion}
In this study, we evaluated the impact of an immersive and interactive VR activity, compared with similarly interactive hands-on and desktop simulations. It is clear there was significant improvement in all conditions in overall test scores from the pre- to post-test, but differences in learning between conditions were not significant (Fig. \ref{fig:Histograms}). Previous work using these assessments found that typical astronomy students score just between 30 and 50\% at pre-test and after a course in astronomy score about 65\% \cite{lindell2004,lindell2005}. This suggests that our participants, from various majors, achieved similar learning gains on the subject of moon phases after a single 10 minute activity as to those obtained in typical instruction. While the non-significant difference between conditions does not indicate virtual reality as one that enhances learning over traditional methods, it does show that VR can perform just as well. The hands-on and desktop activities both used technology very familiar with our participants. If participants can learn the same amount from a new technology they are not familiar with and running a simulation with a low production value there is likely still room for improvement for VR as a teaching tool. 

We can also see how this activity in general performs as a teaching tool for the various Moon phase topic areas. There is clear evidence that further instruction is required on the topics of scale, Moon rotation, reasoning behind phases, and the rise/set times of certain phases (Fig. \ref{fig:prepostbyq}). Students typically struggle with understanding the topic of rise and set times for the Moon phases \cite{lindell2004}. Our results indicate that VR does not help in these areas any better than the traditional methods. Learning in these areas may improve with alternate teaching methods since the results also suggest that the learning activity may be the stronger factor for student learning than the modality of instruction. The population differences between conditions may also be making it difficult to determine the advantages of each condition.
 
As in the previous work, we saw dramatic differences in participants' preference towards the VR simulation. It is possible that VR could be a powerful tool for improving student attitudes towards learning, while obtaining the same learning benefits as other modalities. This impact should be considered as researchers aim to develop ways to recruit more STEM majors or to promote knowledge retention. 

 \section{Future work}
Previous work found no differences in student learning in VR, video, and static images, but did find interactions with gender and video game experience \cite{Smith2017}. We are currently analyzing our data for these effects and how performance relates to previous VR experience. We also aim to further probe students' attitudes towards and experiences in the activities through analysis of the video recordings, time spent in the simulations, and tracking movement. Future publications will provided greater detail on the population, methods, analysis, and results of the work presented in this paper and explore whether the various modalities provide different benefits to different learners. 
\newline
\section{Conclusions}
A cursory glance reveals that VR is as good at teaching Moon phases as traditional methods. The inexperience of participants and the current technical limits on VR appear to be the main limitations in providing a truly immersive experience and distract users from the concepts being taught much as hands-on activities do. Unlike hands-on activities however, these distractions can be overcome by advances in the technology. Higher frame rates, and larger fields of view will eventually remove simulator sickness as an obstacle to immersive interaction with the environment. At its current stage VR is a highly favored way to show the same learning gains as the tried-and-true methods.

\acknowledgements This work was supported by Oculus Education. We would like to thank all the graduate and undergraduate students who helped develop the simulations and run participants.

\bibliography{MoonPhases.bib}

\providecommand{\noopsort}[1]{}\providecommand{\singleletter}[1]{#1}%
\begin{thebibliography}{17}%
\makeatletter
\providecommand \@ifxundefined [1]{%
 \@ifx{#1\undefined}
}%
\providecommand \@ifnum [1]{%
 \ifnum #1\expandafter \@firstoftwo
 \else \expandafter \@secondoftwo
 \fi
}%
\providecommand \@ifx [1]{%
 \ifx #1\expandafter \@firstoftwo
 \else \expandafter \@secondoftwo
 \fi
}%
\providecommand \natexlab [1]{#1}%
\providecommand \enquote  [1]{``#1''}%
\providecommand \bibnamefont  [1]{#1}%
\providecommand \bibfnamefont [1]{#1}%
\providecommand \citenamefont [1]{#1}%
\providecommand \href@noop [0]{\@secondoftwo}%
\providecommand \href [0]{\begingroup \@sanitize@url \@href}%
\providecommand \@href[1]{\@@startlink{#1}\@@href}%
\providecommand \@@href[1]{\endgroup#1\@@endlink}%
\providecommand \@sanitize@url [0]{\catcode `\\12\catcode `\$12\catcode
  `\&12\catcode `\#12\catcode `\^12\catcode `\_12\catcode `\%12\relax}%
\providecommand \@@startlink[1]{}%
\providecommand \@@endlink[0]{}%
\providecommand \url  [0]{\begingroup\@sanitize@url \@url }%
\providecommand \@url [1]{\endgroup\@href {#1}{\urlprefix }}%
\providecommand \urlprefix  [0]{URL }%
\providecommand \Eprint [0]{\href }%
\providecommand \doibase [0]{http://dx.doi.org/}%
\providecommand \selectlanguage [0]{\@gobble}%
\providecommand \bibinfo  [0]{\@secondoftwo}%
\providecommand \bibfield  [0]{\@secondoftwo}%
\providecommand \translation [1]{[#1]}%
\providecommand \BibitemOpen [0]{}%
\providecommand \bibitemStop [0]{}%
\providecommand \bibitemNoStop [0]{.\EOS\space}%
\providecommand \EOS [0]{\spacefactor3000\relax}%
\providecommand \BibitemShut  [1]{\csname bibitem#1\endcsname}%
\let\auto@bib@innerbib\@empty
\bibitem [{\citenamefont {Anderson}(2003)}]{ANDERSON200391}%
  \BibitemOpen
  \bibfield  {author} {\bibinfo {author} {\bibfnamefont {M.~L.}\ \bibnamefont
  {Anderson}},\ }\href {\doibase https://doi.org/10.1016/S0004-3702(03)00054-7}
  {\bibfield  {journal} {\bibinfo  {journal} {Artificial Intelligence}\
  }\textbf {\bibinfo {volume} {149}},\ \bibinfo {pages} {91 } (\bibinfo {year}
  {2003})}\BibitemShut {NoStop}%
\bibitem [{\citenamefont {Finkelstein}\ \emph {et~al.}(2005)\citenamefont
  {Finkelstein}, \citenamefont {Adams}, \citenamefont {Keller}, \citenamefont
  {Kohl}, \citenamefont {Perkins}, \citenamefont {Podolefsky}, \citenamefont
  {Reid},\ and\ \citenamefont {LeMaster}}]{Finkelstein2005}%
  \BibitemOpen
  \bibfield  {author} {\bibinfo {author} {\bibfnamefont {N.~D.}\ \bibnamefont
  {Finkelstein}}, \bibinfo {author} {\bibfnamefont {W.~K.}\ \bibnamefont
  {Adams}}, \bibinfo {author} {\bibfnamefont {C.~J.}\ \bibnamefont {Keller}},
  \bibinfo {author} {\bibfnamefont {P.~B.}\ \bibnamefont {Kohl}}, \bibinfo
  {author} {\bibfnamefont {K.~K.}\ \bibnamefont {Perkins}}, \bibinfo {author}
  {\bibfnamefont {N.~S.}\ \bibnamefont {Podolefsky}}, \bibinfo {author}
  {\bibfnamefont {S.}~\bibnamefont {Reid}}, \ and\ \bibinfo {author}
  {\bibfnamefont {R.}~\bibnamefont {LeMaster}},\ }\href {\doibase
  10.1103/PhysRevSTPER.1.010103} {\bibfield  {journal} {\bibinfo  {journal}
  {Phys. Rev. ST Phys. Educ. Res.}\ }\textbf {\bibinfo {volume} {1}},\ \bibinfo
  {pages} {010103} (\bibinfo {year} {2005})}\BibitemShut {NoStop}%
\bibitem [{\citenamefont {Wieman}(2015)}]{Weiman2015}%
  \BibitemOpen
  \bibfield  {author} {\bibinfo {author} {\bibfnamefont {C.}~\bibnamefont
  {Wieman}},\ }\href {\doibase 10.1119/1.4928349} {\bibfield  {journal}
  {\bibinfo  {journal} {The Physics Teacher}\ }\textbf {\bibinfo {volume}
  {53}},\ \bibinfo {pages} {349} (\bibinfo {year} {2015})}\BibitemShut
  {NoStop}%
\bibitem [{\citenamefont {Adams}\ \emph {et~al.}(2008)\citenamefont {Adams},
  \citenamefont {Paulson},\ and\ \citenamefont {Wieman}}]{adams2008}%
  \BibitemOpen
  \bibfield  {author} {\bibinfo {author} {\bibfnamefont {W.~K.}\ \bibnamefont
  {Adams}}, \bibinfo {author} {\bibfnamefont {A.}~\bibnamefont {Paulson}}, \
  and\ \bibinfo {author} {\bibfnamefont {C.~E.}\ \bibnamefont {Wieman}},\
  }\href {\doibase 10.1063/1.3021273} {\bibfield  {journal} {\bibinfo
  {journal} {AIP Conference Proceedings}\ }\textbf {\bibinfo {volume} {1064}},\
  \bibinfo {pages} {59} (\bibinfo {year} {2008})}\BibitemShut {NoStop}%
\bibitem [{\citenamefont {Bricken}(1991)}]{Bricken:1991:VRL:126640.126657}%
  \BibitemOpen
  \bibfield  {author} {\bibinfo {author} {\bibfnamefont {M.}~\bibnamefont
  {Bricken}},\ }\href {\doibase 10.1145/126640.126657} {\bibfield  {journal}
  {\bibinfo  {journal} {SIGGRAPH Comput. Graph.}\ }\textbf {\bibinfo {volume}
  {25}},\ \bibinfo {pages} {178} (\bibinfo {year} {1991})}\BibitemShut
  {NoStop}%
\bibitem [{\citenamefont {Perone}(2016)}]{Perone2016}%
  \BibitemOpen
  \bibfield  {author} {\bibinfo {author} {\bibfnamefont {B.}~\bibnamefont
  {Perone}},\ }\href {https://purl.stanford.edu/mp832xm1650} {\bibfield
  {journal} {\bibinfo  {journal} {Stanford}\ } (\bibinfo {year}
  {2016})}\BibitemShut {NoStop}%
\bibitem [{\citenamefont {Won}\ \emph {et~al.}(2014)\citenamefont {Won},
  \citenamefont {Bailenson},\ and\ \citenamefont {Janssen}}]{won}%
  \BibitemOpen
  \bibfield  {author} {\bibinfo {author} {\bibfnamefont {A.~S.}\ \bibnamefont
  {Won}}, \bibinfo {author} {\bibfnamefont {J.~N.}\ \bibnamefont {Bailenson}},
  \ and\ \bibinfo {author} {\bibfnamefont {J.~H.}\ \bibnamefont {Janssen}},\
  }\href {\doibase 10.1109/TAFFC.2014.2329304} {\bibfield  {journal} {\bibinfo
  {journal} {IEEE Transactions on Affective Computing}\ }\textbf {\bibinfo
  {volume} {5}},\ \bibinfo {pages} {112} (\bibinfo {year} {2014})}\BibitemShut
  {NoStop}%
\bibitem [{\citenamefont {Smith}\ \emph {et~al.}(2017)\citenamefont {Smith},
  \citenamefont {Byrum}, \citenamefont {McCormick}, \citenamefont {Young},
  \citenamefont {Orban},\ and\ \citenamefont {Porter}}]{Smith2017}%
  \BibitemOpen
  \bibfield  {author} {\bibinfo {author} {\bibfnamefont {J.~R.}\ \bibnamefont
  {Smith}}, \bibinfo {author} {\bibfnamefont {A.}~\bibnamefont {Byrum}},
  \bibinfo {author} {\bibfnamefont {T.~M.}\ \bibnamefont {McCormick}}, \bibinfo
  {author} {\bibfnamefont {N.}~\bibnamefont {Young}}, \bibinfo {author}
  {\bibfnamefont {C.}~\bibnamefont {Orban}}, \ and\ \bibinfo {author}
  {\bibfnamefont {C.~D.}\ \bibnamefont {Porter}},\ }\bibfield  {booktitle}
  {\emph {\bibinfo {booktitle} {Physics Education Research Conference 2017}},\
  }\href@noop {} {\ \bibinfo {series} {PER Conference},\ \bibinfo {pages} {376}
  (\bibinfo {year} {2017})}\BibitemShut {NoStop}%
\bibitem [{\citenamefont {Salehi}\ \emph {et~al.}(2015)\citenamefont {Salehi},
  \citenamefont {Keil}, \citenamefont {Kuo},\ and\ \citenamefont
  {Wieman}}]{Salehi2015}%
  \BibitemOpen
  \bibfield  {author} {\bibinfo {author} {\bibfnamefont {S.}~\bibnamefont
  {Salehi}}, \bibinfo {author} {\bibfnamefont {M.}~\bibnamefont {Keil}},
  \bibinfo {author} {\bibfnamefont {E.}~\bibnamefont {Kuo}}, \ and\ \bibinfo
  {author} {\bibfnamefont {C.}~\bibnamefont {Wieman}},\ }in\ \href@noop {}
  {\emph {\bibinfo {booktitle} {Physics Education Research Conference 2015}}},\
  \bibinfo {series and number} {PER Conference}\ (\bibinfo {address} {College
  Park, MD},\ \bibinfo {year} {2015})\ pp.\ \bibinfo {pages}
  {291--294}\BibitemShut {NoStop}%
\bibitem [{\citenamefont {Prather}(2008)}]{tutorial}%
  \BibitemOpen
  \bibfield  {author} {\bibinfo {author} {\bibfnamefont {E.}~\bibnamefont
  {Prather}},\ }\href@noop {} {\emph {\bibinfo {title} {Lecture tutorials for
  introductory astronomy}}}\ (\bibinfo  {publisher} {Pearson},\ \bibinfo {year}
  {2008})\BibitemShut {NoStop}%
\bibitem [{\citenamefont {Newbury}(2011)}]{newbury}%
  \BibitemOpen
  \bibfield  {author} {\bibinfo {author} {\bibfnamefont {P.}~\bibnamefont
  {Newbury}},\ }\href {https://peternewbury.org/2011/09/06/phases-of-the-moon/}
  {\enquote {\bibinfo {title} {Phases of the moon},}\ } (\bibinfo {year}
  {2011})\BibitemShut {NoStop}%
\bibitem [{\citenamefont {{Hufnagel}}(2002)}]{2002AEdRv...1a..47H}%
  \BibitemOpen
  \bibfield  {author} {\bibinfo {author} {\bibfnamefont {B.}~\bibnamefont
  {{Hufnagel}}},\ }\href@noop {} {\bibfield  {journal} {\bibinfo  {journal}
  {Astronomy Education Review}\ }\textbf {\bibinfo {volume} {1}},\ \bibinfo
  {pages} {47} (\bibinfo {year} {2002})}\BibitemShut {NoStop}%
\bibitem [{\citenamefont {Lindell}\ and\ \citenamefont
  {Olsen}(2002)}]{lindell2002}%
  \BibitemOpen
  \bibfield  {author} {\bibinfo {author} {\bibfnamefont {R.}~\bibnamefont
  {Lindell}}\ and\ \bibinfo {author} {\bibfnamefont {J.~P.}\ \bibnamefont
  {Olsen}},\ }in\ \href@noop {} {\emph {\bibinfo {booktitle} {Physics Education
  Research Conference 2002}}},\ \bibinfo {series and number} {PER Conference}\
  (\bibinfo {address} {Boise, Idaho},\ \bibinfo {year} {2002})\BibitemShut
  {NoStop}%
\bibitem [{\citenamefont {Lindell}()}]{lindell2005}%
  \BibitemOpen
  \bibfield  {author} {\bibinfo {author} {\bibfnamefont {R.}~\bibnamefont
  {Lindell}},\ }in\ \href@noop {} {\emph {\bibinfo {booktitle} {Physics
  Education Research Conference 2004}}},\ pp.\ \bibinfo {pages}
  {53--56}\BibitemShut {NoStop}%
\bibitem [{\citenamefont {Slater}(2014)}]{slater2014}%
  \BibitemOpen
  \bibfield  {author} {\bibinfo {author} {\bibfnamefont {S.~J.}\ \bibnamefont
  {Slater}},\ }\href@noop {} {\bibfield  {journal} {\bibinfo  {journal} {J.
  Astro. Earth. Sci. Educ.}\ }\textbf {\bibinfo {volume} {1}},\ \bibinfo
  {pages} {22} (\bibinfo {year} {2014})}\BibitemShut {NoStop}%
\bibitem [{\citenamefont {Wilhelm}\ \emph {et~al.}(2018)\citenamefont
  {Wilhelm}, \citenamefont {Cole}, \citenamefont {Cohen},\ and\ \citenamefont
  {Lindell}}]{wilhelm2018}%
  \BibitemOpen
  \bibfield  {author} {\bibinfo {author} {\bibfnamefont {J.}~\bibnamefont
  {Wilhelm}}, \bibinfo {author} {\bibfnamefont {M.}~\bibnamefont {Cole}},
  \bibinfo {author} {\bibfnamefont {C.}~\bibnamefont {Cohen}}, \ and\ \bibinfo
  {author} {\bibfnamefont {R.}~\bibnamefont {Lindell}},\ }\href {\doibase
  10.1103/PhysRevPhysEducRes.14.010150} {\bibfield  {journal} {\bibinfo
  {journal} {Phys. Rev. Phys. Educ. Res.}\ }\textbf {\bibinfo {volume} {14}},\
  \bibinfo {pages} {010150} (\bibinfo {year} {2018})}\BibitemShut {NoStop}%
\bibitem [{\citenamefont {Lindell}\ and\ \citenamefont
  {Sommer}(2004)}]{lindell2004}%
  \BibitemOpen
  \bibfield  {author} {\bibinfo {author} {\bibfnamefont {R.~S.}\ \bibnamefont
  {Lindell}}\ and\ \bibinfo {author} {\bibfnamefont {S.~R.}\ \bibnamefont
  {Sommer}},\ }\href {\doibase 10.1063/1.1807257} {\bibfield  {journal}
  {\bibinfo  {journal} {AIP Conference Proceedings}\ }\textbf {\bibinfo
  {volume} {720}},\ \bibinfo {pages} {73} (\bibinfo {year} {2004})}\BibitemShut
  {NoStop}%
\end{thebibliography}%

\end{document}